\input epsf
\documentstyle[aps]{revtex}
\tighten
\begin{document}


\preprint{HUTP-02/A009,hep-th/0204xxx}

\title{Brane cosmology driven by the rolling tachyon}
\author{Shinji Mukohyama}
\address{
Department of Physics, Harvard University\\
Cambridge, MA, 02138, USA
}
\date{\today}

\maketitle

\begin{abstract} 
 Brane cosmology driven by the tachyon rolling down to its ground
 state is investigated. We adopt an effective field theoretical
 description for the tachyon and Randall-Sundrum type brane world
 scenario. After formulating basic equations, we show that the standard
 cosmology with a usual scalar field can mimic the low energy behavior
 of the system near the tachyon ground state. We also investigate
 qualitative behavior of the system beyond the low energy regime for
 positive, negative and vanishing $4$-dimensional effective cosmological
 constant $\Lambda_4=\kappa_5^4V(T_0)^2/12-|\Lambda_5|/2$, where
 $\kappa_5$ and $\Lambda_5$ are $5$-dimensional gravitational coupling
 constant and (negative) cosmological constant, respectively, and
 $V(T_0)$ is the (positive) tension of the brane in the tachyon ground
 state. In particular, for $\Lambda_4<0$ the tachyon never settles down
 to its potential minimum and the universe eventually hits a big-crunch 
 singularity. 
\end{abstract}


\section{Introduction}


Pioneered by Sen~\cite{Sen}, the study of non-BPS objects such as
non-BPS branes, brane-antibrane configurations or spacelike
branes~\cite{sbrane} has been attracting physical interests in string 
theory~\cite{review}. These objects are expected to be important for our
understanding non-perturbative dualities beyond the BPS level. Moreover,
they may play roles in cosmology. Sen showed that classical decay of
unstable D-brane in string theories produces pressure-less gas with
non-zero energy density~\cite{Sen2}. Gibbons took into account coupling
to gravitational field by adding an Einstein-Hilbert term to the
effective action of the tachyon on a brane, and initiated a study of
``tachyon cosmology'', cosmology driven by the tachyon rolling down to
its ground state~\cite{Gibbons}. Fairbairn and Tytgat considered
possibility of inflation driven by the rolling tachyon~\cite{FT}.


Another subject in which branes play important roles is the brane-world
scenario. Actually, in this scenario a brane in a higher dimensional
bulk spacetime is supposed to be our universe itself. Randall and
Sundrum~\cite{RS} showed that, in a $5$-dimensional AdS background,
$4$-dimensional Newton's law of gravity can be reproduced on the
world-volume of a $3$-brane, despite the existence of the infinite fifth
dimension. There are also cosmological solutions in this scenario, in
which the standard cosmology is restored at low energy, provided that a
parameter representing the mass of a bulk black-hole is small 
enough~\cite{CGS,FTW,BDEL,Mukohyama2000a,Kraus,Ida}. 


There are several papers in which effects of a tachyon is discussed in
the context of brane world scenarios. For example, Papantonopoulos and
Pappa~\cite{PP} discussed a brane world scenario with a tachyon in the
bulk. Alexander~\cite{Alexander} and Mazumdar et. al.~\cite{MPP}
considered inflation on a brane caused by higher-dimensional
$D$-$\bar{D}$ brane annihilation.


Considering that the tachyon is a degree of freedom on a brane, 
it is perhaps interesting to consider the brane-world scenario to 
take into account gravity in the tachyon cosmology. In this case, the
Einstein-Hilbert term is introduced in the bulk action rather than in
the brane action. Hence, purpose of this paper is to initiate the
brane-world version of the rolling tachyon cosmology. We adopt an
effective field theoretical description for the tachyon and
Randall-Sundrum type brane world scenario. After formulating basic
equations, we show that the standard cosmology with a usual scalar field
can mimic the low energy behavior of the system near the tachyon ground
state. An obvious consequence of this result is that if the tachyon
potential has a minimum at finite distance from the maximum and if it
oscillates about the minimum then the tachyon will behave like
pressure-less gas. We also investigate qualitative behavior of the
system beyond the low energy regime for positive, negative and vanishing
$4$-dimensional effective cosmological constant.


The rest of this paper is organized as follows. In
Sec.~\ref{sec:tachyon} we briefly review the effective field theoretical
description for the rolling tachyon. In Sec.~\ref{sec:brane} we consider
the Randall-Sundrum type brane cosmology driven by the rolling tachyon
and investigate the low energy behavior of the system near the tachyon
ground state. In Sec.~\ref{sec:example} we investigate a simple
double-well potential and a run-away potential to see qualitative
behavior of the system beyond the vicinity of the tachyon ground
state. Sec.~\ref{sec:summary} is devoted to a summary of this paper.


\section{Rolling tachyon}
\label{sec:tachyon}

Let us consider an $n$-brane in $D$-dimensional spacetime. The imbedding
of the world volume of the brane can be described by the parametric
equation 
%
\begin{equation}
 x^M = Z^M(y),
\end{equation}
where $\{x^M\}$ is a coordinate system in the $D$-dimensional bulk and
$y$ denotes a set of $(n+1)$ parameters $\{y^{\mu}\}$. The parameters 
$y^{\mu}$ will play a role of $(n+1)$-dimensional coordinates on the
world-volume of the $n$-brane. Throughout this paper we shall adopt the
effective field theoretical description of non-BPS branes proposed in
refs.~\cite{Sen2,Garousi,BRWEP,Kluson}, which includes a tachyon. The 
effective action for the brane with the tachyon field $T$ is 
%
\begin{equation}
 S_{brane} = -\int d^{n+1}y \sqrt{|\det\tilde{q}|}V(T), 
  \label{eqn:tachyon-action}
\end{equation}
where $V(T)$ is a tachyon potential, 
%
\begin{equation}
 \tilde{q}_{\mu\nu} = q_{\mu\nu}+\partial_{\mu}T\partial_{\nu}T,
\end{equation}
$q_{\mu\nu}$ is the induced metric defined by
%
\begin{equation}
 q_{\mu\nu} = g_{MN}\frac{\partial Z^M}{\partial y^{\mu}}
  \frac{\partial Z^N}{\partial y^{\nu}},
\end{equation}
and $g_{MN}$ is the bulk metric.

Note that the choice of the tachyon potential $V(T)$ is very
important. Different potentials give different dynamics of the tachyon
and the brane universe. Hence, if we could, we would like to use
potentials predicted by rigorous calculations based on boundary string
field theory or conformal field theory for configurations of our
interest. As we shall see, we would like to consider potentials with a
positive value at the tachyon ground state. This situation may be
expected for, for example, decay of stacked $N$ $D$-branes and $\bar{N}$
$\bar{D}$-branes with $N\ne\bar{N}$. However, as far as the author
knows, there is no such calculation in the literature so far. Hence, in
this paper we shall take an alternative approach: we shall consider
very simple forms of the tachyon potential. In Sec.~\ref{sec:example} we
shall consider a simple double-well potential $V(T)=(T^2-1)^2+V_0$ and a
simple run-away potential $V(T)=1/\cosh(T)+V_0$ to investigate
qualitative behavior of the system.

The surface stress energy tensor $S^{\mu\nu}$ is given by 
%
\begin{equation}
 S^{\mu\nu} \equiv 
  \frac{2}{\sqrt{|q|}}\frac{\delta S_{brane}}{\delta q_{\mu\nu}} 
 = -\sqrt{\left|\frac{\det\tilde{q}}{\det q}\right|}V(T)
  (\tilde{q}^{-1})^{\mu\nu}. 
\end{equation}
The equation of motion for the tachyon is 
%
\begin{equation}
 \frac{1}{\sqrt{|\det\tilde{q}|}}\partial_{\mu}
  \left[\sqrt{|\det\tilde{q}}V(T)(\tilde{q}^{-1})^{\mu\nu}
   \partial_{\nu}T\right] -V'(T) = 0.
\end{equation}

For a homogeneous isotropic brane, we can assume the following form of
the induced metric $q_{\mu\nu}$ and the tachyon field $T$ without loss
of generality.
%
\begin{eqnarray}
 q_{\mu\nu}dy^{\mu}dy^{\nu} & = & -dt^2 + a(t)^2\Omega^K_{ij}dy^idy^j, 
  \nonumber\\
 T & = & T(t),
\end{eqnarray}
where $\Omega^K_{ij}$ is the metric of the $n$-dimensional constant
curvature space with the curvature constant $K$:
%
\begin{equation}
 \Omega^K_{ij}dy^idy^j = \frac{d\rho^2}{1-K\rho^2}+\rho^2d\Omega_{n-1}^2.
\end{equation}
Here, $d\Omega_{n-1}^2$ is the metric of the $(n-1)$-dimensional unit
sphere. Positive, zero and negative values of $K$ correspond to
$S^n$, $R^n$ and $H^n$, respectively. 
With the above form of $q_{\mu\nu}$ and $T$, the equation of motion and
the surface stress energy tensor are reduced to
%
\begin{equation}
 \frac{V\ddot{T}}{1-\dot{T}^2}+n\frac{\dot{a}}{a}V\dot{T}+V'=0,
  \label{eqn:eom}
\end{equation}
%
\begin{equation}
 S^{\mu}_{\nu} = \left(\begin{array}{cccc}
        -\rho & 0 & 0 & 0 \\
        0 & p & 0 & 0 \\
        0 & 0 & \ddots & 0 \\
        0 & 0 & 0 & p 
        \end{array}\right),
\end{equation}
where a dot denotes derivative with respect to $t$, and 
%
\begin{eqnarray}
 \rho & = & \frac{V}{\sqrt{1-\dot{T}^2}}, \nonumber\\
 p & = & -V\sqrt{1-\dot{T}^2}.
  \label{eqn:rho-p}
\end{eqnarray}
The equation of motion is formally equivalent to
the conservation equation $\nabla_{\mu}S^{\mu}_{\nu}=0$, or
%
\begin{equation}
 \dot{\rho}+n\frac{\dot{a}}{a}(\rho+p)=0.
  \label{eqn:conservation-eq}
\end{equation}


\section{Brane cosmology}
\label{sec:brane}

Now we consider brane cosmology driven by the rolling tachyon. We
consider Randall-Sundrum brane world scenario on a $3$-brane ($n=3$) in
a $5$-dimensional bulk spacetime ($D=5$)~\cite{RS}. We assume that the
bulk is invariant under the $Z_2$ reflection along the brane and
described by $5$-dimensional Einstein pure gravity with a negative
cosmological constant and that the brane motion is described by Israel's
junction condition~\cite{Israel}. With these assumptions, the bulk
geometry is AdS-Schwarzschild spacetime~\cite{MSM} and the evolution of
the brane is governed by~\cite{BDL,CGKT}
%
\begin{equation}
 \left(\frac{\dot{a}}{a}\right)^2 = 
  \frac{\kappa_5^4}{36}\rho^2 - \frac{K}{a^2} 
  + \frac{\mu}{a^4} -\frac{1}{l^2}, \label{eqn:gen-Friedmann}
\end{equation}
where $\kappa_5$ is the $5$-dimensional gravitational coupling constant,
$l=\sqrt{6/|\Lambda_5|}$ is the length scale of the bulk (negative)
cosmological constant $\Lambda_5$, and $\mu$ is the mass parameter of
the bulk black hole. This equation looks very different from the
standard Friedmann equation in the sense that the first term in the
right hand side is proportional to $\rho^2$. Nonetheless, the standard
cosmology is restored at low energy if brane tension is properly 
introduced~\cite{CGS,FTW,BDEL,Mukohyama2000a,Kraus,Ida}. The term
$\mu/a^4$ is due to Weyl tensor in the bulk~\cite{SMS} and can be
understood as dark radiation~\cite{Mukohyama2000a}. Thus, our basic
equations for brane cosmology driven by the rolling tachyon are the
equation of motion (\ref{eqn:eom}) and the generalized Friedmann
equation (\ref{eqn:gen-Friedmann}) with $\rho$ given by
(\ref{eqn:rho-p}).

Let us analyze behavior of the system near the tachyon ground state
$T=T_0$ and show that the standard cosmology with a usual scalar field
can mimic the low energy behavior of the system. For this purpose, we
assume that 
%
\begin{equation}
 [V(T)-V(T_0)]/V(T_0)
  \sim lV'(T)/V(T_0)
  \sim \dot{T}^2
  \sim l\ddot{T}
  \sim O(\epsilon),\label{eqn:assumption}
\end{equation}
where $\epsilon$ is a dimensionless small parameter. The actual value
of $T_0$ can be either finite or infinite. With this assumption,
the generalized Friedmann equation (\ref{eqn:gen-Friedmann}) and the
tachyon equation (\ref{eqn:eom}) are reduced to 
%
\begin{equation}
 \left(\frac{\dot{a}}{a}\right)^2 =  
  \frac{8\pi G_N}{3}\left[\frac{1}{2}\dot{\phi}^2+V_{eff}(\phi)\right] 
  -\frac{K}{a^2}  + \frac{\Lambda_4}{3} + \frac{\mu}{a^4} + O(\epsilon^2), 
\end{equation}
%
\begin{equation}
 \ddot{\phi}+3\frac{\dot{a}}{a}\dot{\phi}+V_{eff}'(\phi) = O(\epsilon^2),
\end{equation}
where $\phi=\sqrt{V(T_0)}T$, $V_{eff}(\phi)=V(T)-V(T_0)$,
$G_{N}=\kappa_5^4V(T_0)/48\pi$ and
$\Lambda_4=\kappa_5^4V(T_0)^2/12-3l^{-2}$. 
These equations are the same as the corresponding equations in the
standard cosmology driven by a usual scalar field $\phi$ with the
potential $V_{eff}(\phi)$, the cosmological constant $\Lambda_4$ and the
dark radiation $\mu/a^4$, up to corrections of order
$O(\epsilon^2)$. Consistency of the above reduced equations with the
assumption (\ref{eqn:assumption}) requires that 
%
\begin{equation}
 Kl^2a^{-2} \sim l^2\Lambda_4 \sim \mu l^2a^{-4} \sim O(\epsilon). 
  \label{eqn:consistency}
\end{equation}
In particular, we need to impose a fine-tuning between the tachyon
vacuum energy $V(T_0)$ and the bulk cosmological constant $\Lambda_5$ so
that the $4$-dimensional effective cosmological constant $\Lambda_4$ is
small compared to $\Lambda_5$.

It is easy to introduce other matter fields on the brane, following 
refs.~\cite{CGS,FTW,BDEL,Mukohyama2000a,Kraus,Ida}. In this case, the
generalized Friedmann equation (\ref{eqn:gen-Friedmann}) becomes 
%
\begin{equation}
 \left(\frac{\dot{a}}{a}\right)^2 =  
  \frac{8\pi G_N}{3}\left[\frac{1}{2}\dot{\phi}^2+V_{eff}(\phi)
		    + \rho_{matter}\right] 
  -\frac{K}{a^2} + \frac{\Lambda_4}{3} + \frac{\mu}{a^4} + O(\epsilon^2), 
\end{equation}
where $\rho_{matter}$ is energy density of other matter fields on the
brane, and the equation of motion of the tachyon is unchanged. Here, we
have assumed that $\rho_{matter}/V(T_0)=O(\epsilon)$. Therefore, 
the standard cosmology can mimic the low-energy behavior of the brane
cosmology driven by the tachyon.

It is well-known in the standard cosmology that a scalar field
oscillating about a potential minimum behaves like pressure-less 
gas. Hence, if the tachyon potential in the effective field theory has a
minimum at finite distance from the maximum and if it oscillates about
the minimum then the tachyon will behave like pressure-less gas. 
Note that calculations in boundary string field theory suggest that the
minimum of the tachyon potential is a finite distance away from the
maximum~\cite{GS,KMM}. It is probably worth while mentioning Sen's
result that decay of unstable D-branes in string theory produces
pressure-less gas with non-zero energy density but that the minima must
be at infinity in the effective field theory
description~\cite{Sen2}. There is an interesting accidental coincidence
(production of pressure-less gas) between Sen's result in conformal
field theory and the above conclusion based on the effective field
theory and the brane world.


\section{Simple examples}
\label{sec:example}

To see behavior of the system beyond the low energy regime satisfying
(\ref{eqn:assumption}) and (\ref{eqn:consistency}), we need to analyze
the generalized Friedmann equation (\ref{eqn:gen-Friedmann}) and the
tachyon equation (\ref{eqn:eom}) directly. In the following, for
simplicity we consider a spatially flat brane ($K=0$) in the pure AdS
bulk ($\mu=0$). By introducing dimensionless quantities $\tau=t/l$, 
$x=T/l$, $y=\partial_{\tau}x$ and $z=\partial_{\tau}a/a$, our basic
equations in this case are written as  
%
\begin{eqnarray}
 \partial_{\tau}x & = & y, \nonumber\\
 \partial_{\tau}y & = & 
  -(1-y^2)\left[3yz+\frac{v'(x)}{v(x)}\right],\nonumber\\
 \label{eqn:dx-dy}
\end{eqnarray}
where $v(x)=\kappa_5^2lV(T)/6$, $v'(x)=\partial_xv(x)$, and $z$ is given
by 
%
\begin{equation}
 z^2 = \frac{v(x)^2}{1-y^2}-1.
\end{equation}

There are the expanding branch ($z>0$) and the contracting branch
($z<0$). Hence, the projection to the $xy$-plane is two-fold. 
Hereafter, we shall consider the following three cases separately: (i)
$v(x_0)>1$ ($\Lambda_4>0$), where $x_0$ is a potential minimum; (ii) 
$v(x_0)<1$ ($\Lambda_4<0$); (iii) $v(x_0)=1$ ($\Lambda_4=0$). In the
case (i), expanding and contracting branches are disconnected. In the
case (ii) the two branches are connected in the $xyz$-space through the
intersection of the surfaces $v(x)^2=1-y^2$ and $z=0$. In the critical
case (iii), two branches are just touching at a point
$(x,y,z)=(x_0,0,0)$ in the $xyz$-space. (See ref.~\cite{FFKL} for
discussion about the standard scalar field cosmology with negative 
potentials.)

Hereafter, unless otherwise stated, we shall consider the expanding
branch. The behavior of the system can be easily understood by 
plotting the vector field $(\partial_{\tau}x,\partial_{\tau}y)$ in the
$xy$-plane.

\begin{figure}[b]
 \centering\leavevmode\epsfysize=8cm \epsfbox{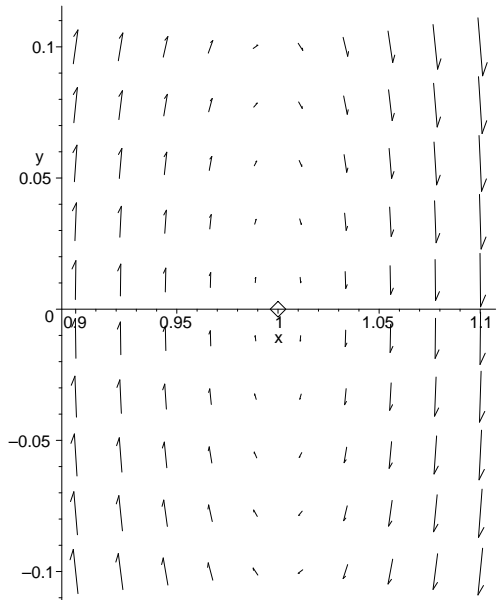}
 \caption{\label{fig:doublewell}A plot of the vector field
 ($\partial_{\tau}x$,  $\partial_{\tau}y$) for the double-well
 potential $v(x)=(x^2-1)^2+1$. The diamond at $(x,y)=(1,0)$ represents
 an attractor corresponding to Minkowski spacetime with the tachyon in its
 ground state.} 
\end{figure}

\begin{figure}[b]
 \centering\leavevmode\epsfysize=8cm \epsfbox{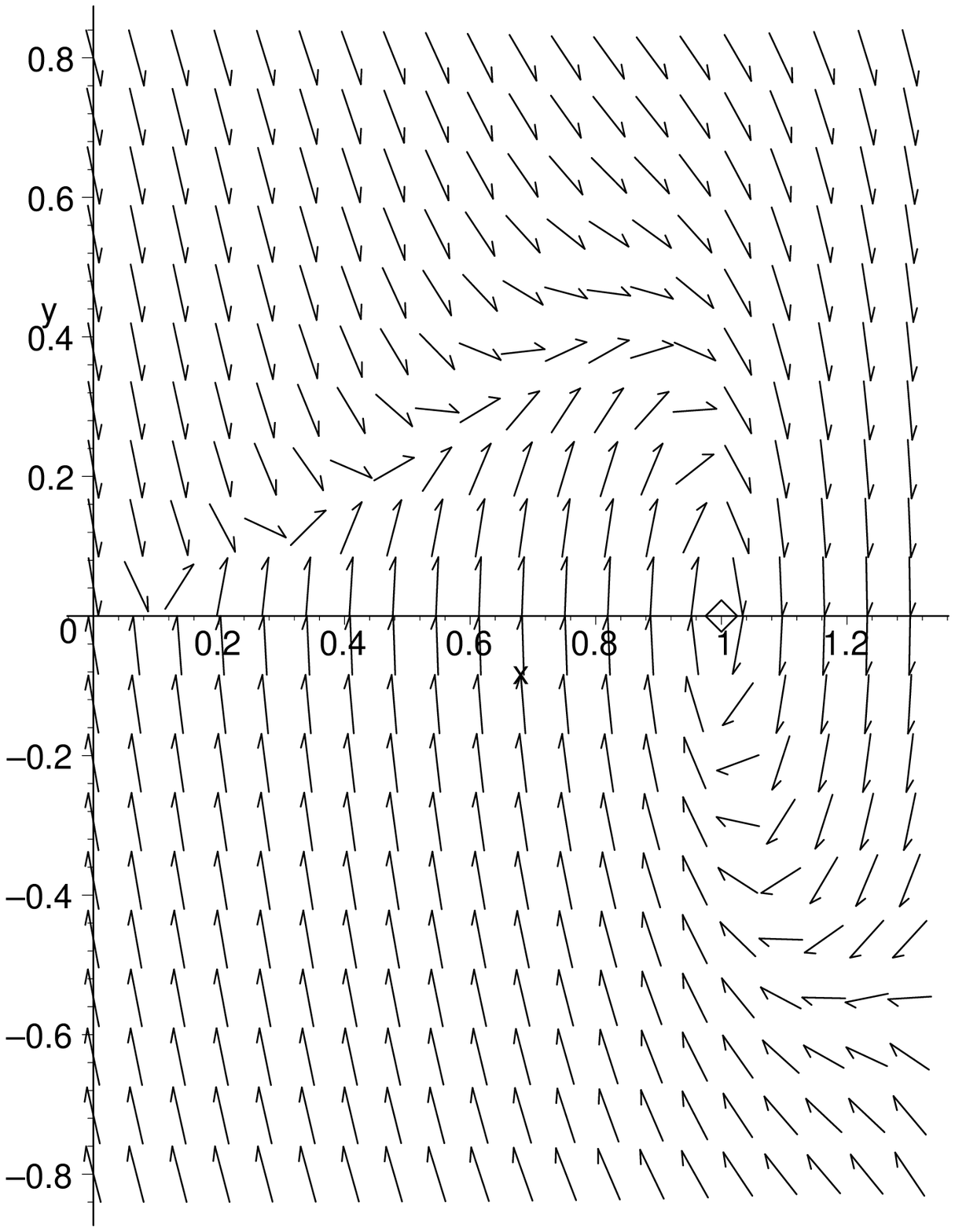}
 \caption{\label{fig:doublewell-norm}The global picture of the
 normalized vector field for the double-well potential
 $v(x)=(x^2-1)^2+1$. The diamond at $(x,y)=(1,0)$ represents
 an attractor corresponding to Minkowski spacetime with the tachyon in its
 ground state.} 
\end{figure}

First, let us consider the critical case (iii). For a simple double-well
potential $v(x)=(x^2-1)^2+1$, figure~\ref{fig:doublewell} shows a plot
near the minimum $x=1$. This plot shows that the tachyon field
oscillates about the minimum. Since the conservation equation
(\ref{eqn:conservation-eq}) implies that expansion of the brane universe
dilutes the energy density $\rho$, the amplitude of the oscillation
decays. In order to see a global picture of the vector field, it is
convenient to introduce a normalized vector field ($N\partial_{\tau}x$, 
$N\partial_{\tau}y$), where
$N=1/\sqrt{(\partial_{\tau}x)^2+(\partial_{\tau}y)^2}$. 
Figure~\ref{fig:doublewell-norm} shows the global picture of the 
normalized vector field. It is easy to see how the tachyon decays into
the minimum with oscillation. Hence, for the critical case (iii), we may 
generally expect that the tachyon rolls down to the minimum of the
potential, that it oscillates about the minimum with decaying amplitude
and that the universe approaches to Minkowski spacetime ($H=z/l=0$).

\begin{figure}[b]
 \centering\leavevmode\epsfysize=8cm \epsfbox{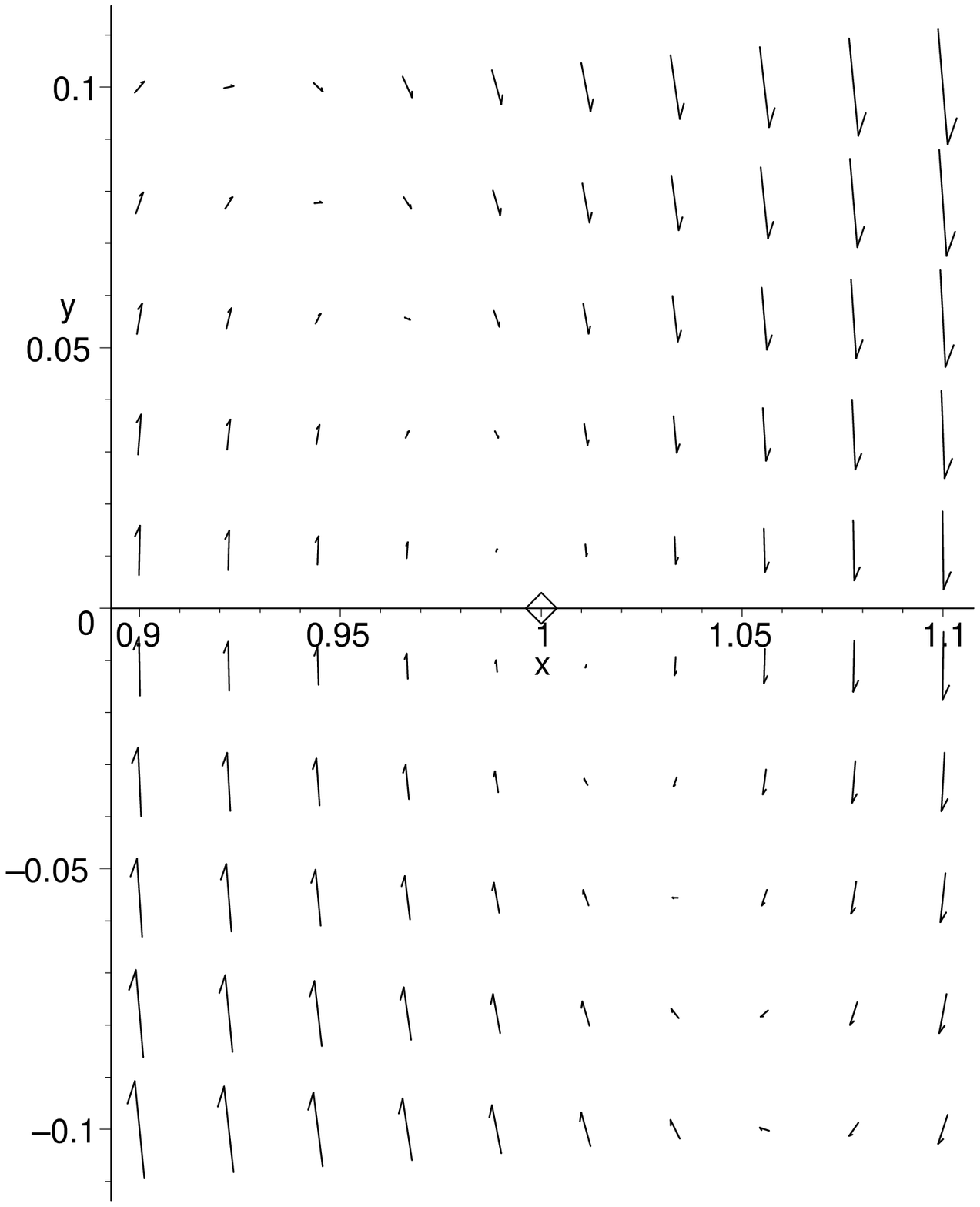}
 \caption{\label{fig:doublewell2}A plot of the vector field
 ($\partial_{\tau}x$,  $\partial_{\tau}y$) for the double-well
 potential $v(x)=(x^2-1)^2+1.5$. The diamond at $(x,y)=(1,0)$ represents
 an attractor corresponding to de Sitter spacetime with the tachyon in
 its ground state.} 
\end{figure}

\begin{figure}[b]
 \centering\leavevmode\epsfysize=8cm \epsfbox{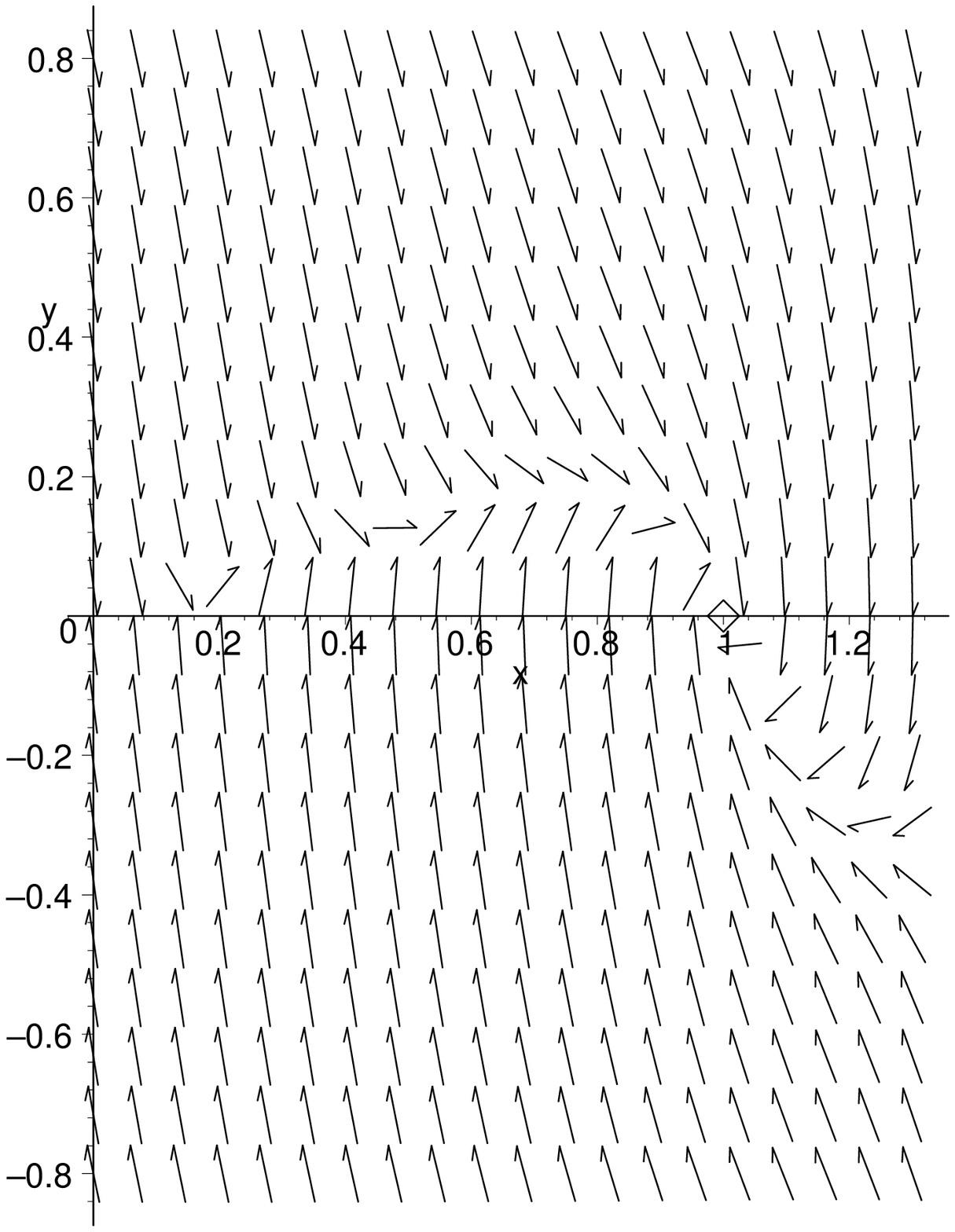}
 \caption{\label{fig:doublewell2-norm}The global picture of the
 normalized vector field for the double-well potential
 $v(x)=(x^2-1)^2+1.5$. The diamond at $(x,y)=(1,0)$ represents
 an attractor corresponding to de Sitter spacetime with the tachyon in
 its ground state.}
\end{figure}

Next, let us consider the case (i). Figure~\ref{fig:doublewell2} and
figure~\ref{fig:doublewell2-norm} show plots for the potential
$v(x)=(x^2-1)^2+1.5$. In this case, rolling down of the tachyon is
slower than the case (iii) due to the larger ``cosmic friction term''
$3yz$ in (\ref{eqn:dx-dy}). For the same reason, decay of the
oscillation about the minimum in the case (i) is faster than the case
(iii). In the case (i) the universe approaches to de Sitter spacetime
($H=z/l>0$).

\begin{figure}[b]
 \centering\leavevmode\epsfysize=8cm \epsfbox{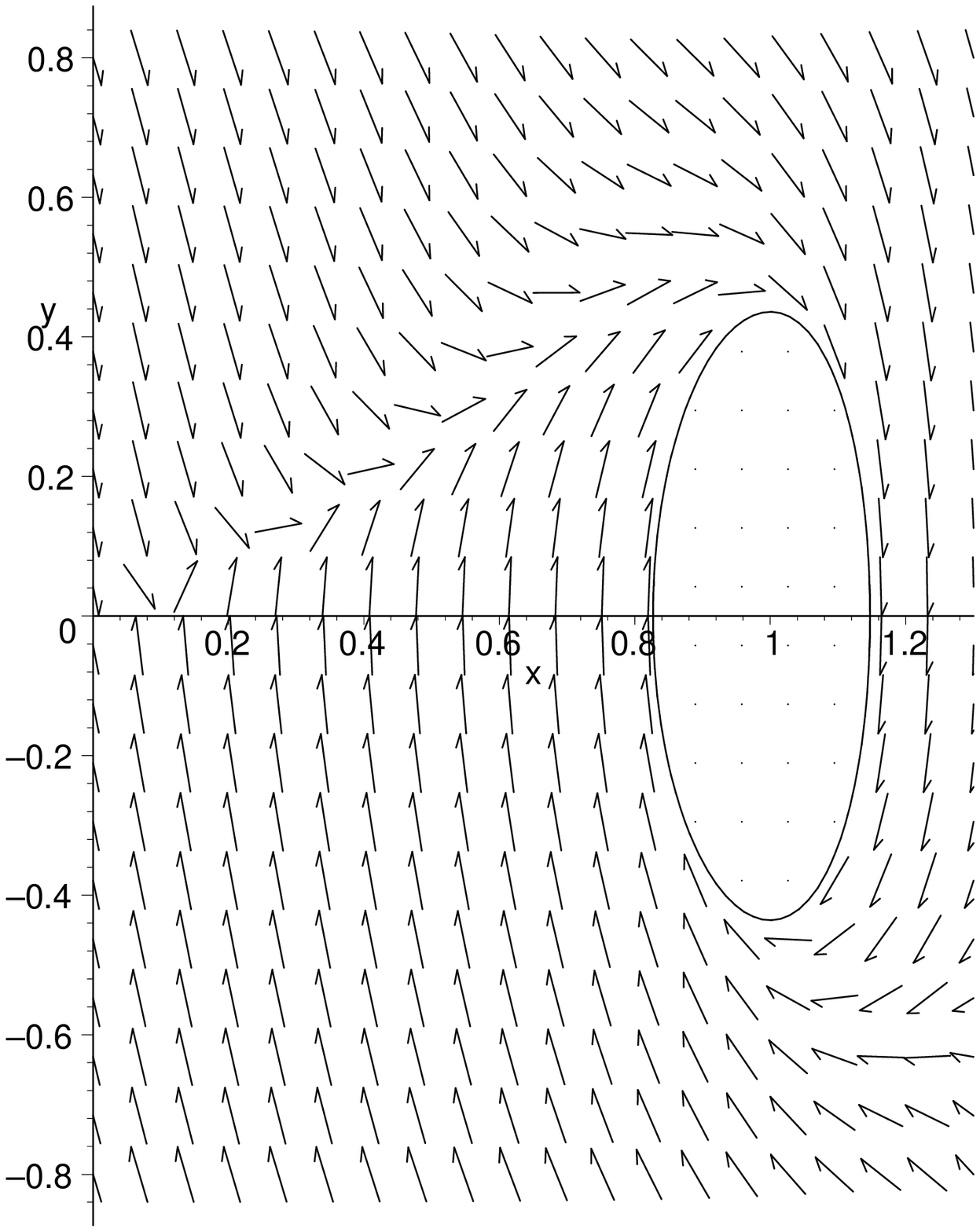}
 \caption{\label{fig:negative-expand}The global picture of the
 normalized vector field for the double-well potential
 $v(x)=(x^2-1)^2+0.9$ in the expanding branch. The region without arrows
 is not allowed. The boundary between the allowed and disallowed regions
 corresponds to the throat through which the system evolves to the
 contracting branch shown in figure~\ref{fig:negative-contract}.}
\end{figure}

\begin{figure}[b]
 \centering\leavevmode\epsfysize=8cm \epsfbox{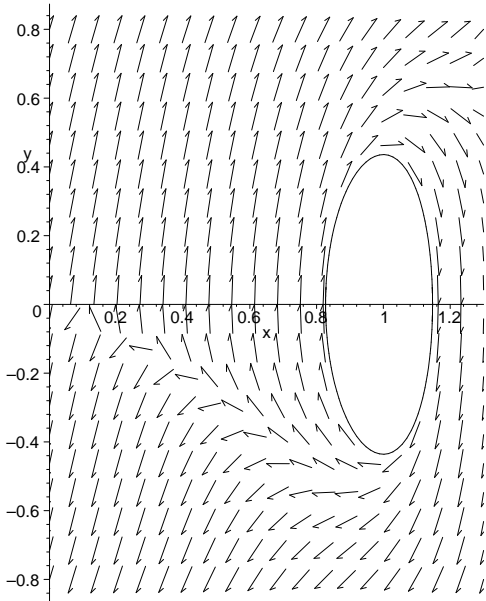}
 \caption{\label{fig:negative-contract}The global picture of the
 normalized vector field for the double-well potential
 $v(x)=(x^2-1)^2+0.9$ in the contracting branch. The region without
 arrows is not allowed.The boundary between the allowed and disallowed
 regions corresponds to the throat through which the system evolves from
 the expanding branch shown in figure~\ref{fig:negative-expand} to the
 contracting branch shown here.} 
\end{figure}

Finally, let us consider the case (ii). Figure~\ref{fig:negative-expand}
and figure~\ref{fig:negative-contract} show the expanding branch and the
contracting branch, respectively, for the potential
$v(x)=(x^2-1)^2+0.9$. These two branches are connected in the
$xyz$-space through a throat. In each figure, the throat is the boundary
between regions with and without arrows. Hence, even if the universe was
initially in the expanding branch, the system approaches to the throat
and goes into the contracting branch. Eventually, the universe hits a
big-crunch singularity. From the previous results of the cases (iii) and
(i), one might have expected that the universe would approach to 
anti-de Sitter spacetime with the tachyon at its potential minimum. As
we already saw, the figures~\ref{fig:negative-expand} and
\ref{fig:negative-contract} show that this is not the case. Actually,
the tachyon never settles down to the potential minimum in the case 
(ii).

The general qualitative behavior explained here is consistent with the
result of ref.~\cite{FFKL} in which the standard scalar field cosmology
was analyzed, provided that $V_0$ in ref.~\cite{FFKL} is replaced by the
$4$-dimensional effective cosmological constant
$\Lambda_4=\kappa_5^4V(T_0)^2/12-|\Lambda_5|/2$.


The above examples of double-well potentials have minima at finite
$T$. The finiteness seems consistent with the result in boundary 
string field theory that the minimum of the tachyon potential is a
finite distance away from the maximum~\cite{GS,KMM}. However, this is
not, at least apparently, consistent with the result in conformal field
theory that the tachyon evolves to the minimum without
oscillation~\cite{Sen1,Sen2}. Sen suggested that the minima must be at
infinity in the effective field theory description~\cite{Sen2}. Hence,
it may be relevant to consider not a double-well potential but a
run-away potential.

\begin{figure}[b]
 \centering\leavevmode\epsfysize=8cm \epsfbox{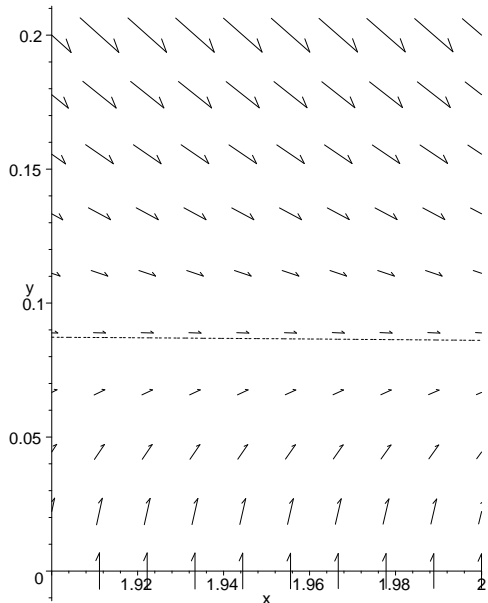}
 \caption{\label{fig:coshinv}A plot of the vector field
 ($\partial_{\tau}x$,  $\partial_{\tau}y$) for the ran-away potential 
 $v(x)=1/\cosh(x)+1$. The dotted line represents points satisfying
 $\partial_{\tau}y=0$.} 
\end{figure}

\begin{figure}[b]
 \centering\leavevmode\epsfysize=8cm \epsfbox{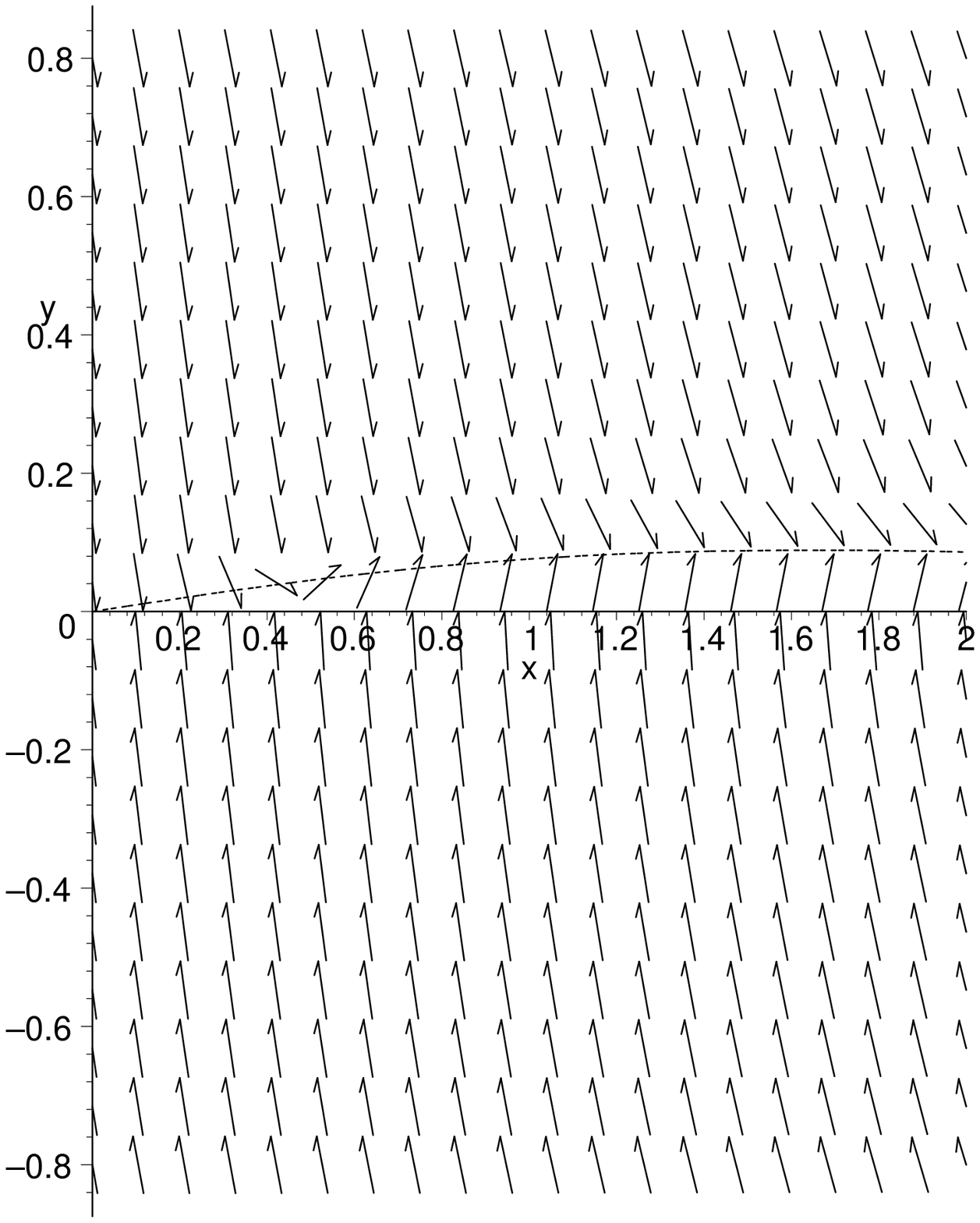}
 \caption{\label{fig:coshinv-norm}The global picture of the normalized
 vector field for the ran-away potential $v(x)=1/\cosh(x)+1$. The dotted
 line represents points satisfying $\partial_{\tau}y=0$.} 
\end{figure}

As a simple example, let us consider a ran-away potential
$v(x)=1/\cosh(x)+1$. Figure~\ref{fig:coshinv} shows a plot of the vector
($\partial_{\tau}x$,$\partial_{\tau}y$). In order to see a global
behavior of the system, figure~\ref{fig:coshinv-norm} plots the
normalized vector field ($N\partial_{\tau}x$, $N\partial_{\tau}y$),
where $N=1/\sqrt{(\partial_{\tau}x)^2+(\partial_{\tau}y)^2}$. From these
two figures, it is easy to see how rolling down of the tachyon slows
down. Hence, we may expect inflation on the brane by the rolling
tachyon, depending on the form of the potential. Actually, the effective
potential $V_{eff}(\phi)$ corresponding to the run-away potential
$v(x)=1/\cosh(x)+1$ is
$V_{eff}(\phi)=6\kappa_4^{-2}l^{-2}/\cosh(\kappa_4\phi/\sqrt{6})$ 
$\sim\exp(-\kappa_4\phi/\sqrt{6})$ ($\kappa_4\phi\gg 1$), where 
$\kappa_4^2=\kappa_5^2/l$, and it is known that in the standard
cosmology the so called power-law inflation occurs for this potential
$V_{eff}(\phi)$~\cite{LM,Halliwell,KM}.



\section{Summary}
\label{sec:summary}

We have investigated brane cosmology driven by the tachyon rolling down
to its ground state. We have adopted the effective action
(\ref{eqn:tachyon-action}) for the tachyon and Randall-Sundrum type
brane world scenario. We have shown that the standard cosmology with a 
usual scalar field can mimic the low energy behavior of the system near
the tachyon ground state as far as the conditions (\ref{eqn:assumption})
and (\ref{eqn:consistency}) are satisfied. In particular, if the tachyon
potential in the effective field theory has a minimum at finite distance
from the maximum and if it oscillates about the minimum then the tachyon
will behave like pressure-less gas. There is an interesting accidental
coincidence (production of pressure-less gas) between Sen's result in
conformal field theory~\cite{Sen1,Sen2} and the above conclusion based
on the effective field theory and the brane world.

We have also analyzed qualitative behavior of the system beyond the
low energy regime for a spatially flat brane in pure AdS bulk. As an
example we have considered a double-well potential. For the double-well
potential, qualitative behavior of the system is classified by the sign
of $\Lambda_4=\kappa_5^4V(T_0)^2/12-|\Lambda_5|/2$: (i) $\Lambda_4>0$;
(ii) $\Lambda_4<0$; (iii) $\Lambda_4=0$. In the case (i) the tachyon
rolls down the potential hill and starts oscillating about a
minimum. The amplitude of the oscillation decays due to cosmic expansion
and the universe approaches to de Sitter spacetime. The case (iii) is
similar to the case (i), but the universe approaches to Minkowski
spacetime. In the case (ii), even if the universe was initially
expanding, it starts contracting and eventually hits a big-crunch
singularity. In this case the tachyon never settles down to the
potential minimum. Finally, we considered a run-away potential, for
which the ground state is at infinity. For the run-away potential with
$\Lambda_4=0$, rolling down of the tachyon slows down and power-law
inflation can occur.

For future works, it is important to investigate qualitative differences
among the following four simple cosmologies: (a1) the standard cosmology 
driven by a usual canonical scalar field; (a2) the standard cosmology
driven by the rolling tachyon; (b1) the brane cosmology driven by
a usual canonical scalar field; (b2) the brane cosmology driven by the
rolling tachyon. In particular, properties of inflation in the early
universe can be very different in these scenarios. It is also
interesting to investigate how these scenarios become indistinguishable
from each other as the system evolves to the low energy regime.

\begin{acknowledgments}
 The author would like to thank Lev Kofman, Shiraz Minwalla, Lisa
 Randall, Ricardo Schiappa and Andrew Strominger for useful discussions 
 and/or comments. He would be grateful to Werner Israel for continuing
 encouragement. This work is supported by JSPS Postdoctoral Fellowship
 for Research Abroad. 
\end{acknowledgments}


\end{document}